\documentclass{article}
\usepackage{arxiv}

\usepackage{graphicx}
\usepackage{authblk}
\usepackage{natbib}
\usepackage{amsmath}
\usepackage{diagbox}
\usepackage{starfont}
\usepackage{amssymb}
\usepackage{hyperref}
\usepackage[final]{changes}%

\graphicspath{{figs/}}

\begin{document}


\title{A study on station-keeping over irregularly shaped asteroids with different sized solar sails}

\author[1]{Lucas Meireles}
\author[1]{Othon Winter}
\author[2]{Antônio Prado}
\affil[1]{São Paulo State University, Guaratinguetá, Brazil}
\affil[2]{National Institute for Space Research, São José dos Campos, Brazil}

\date{}
\maketitle

\begin{abstract}
Asteroid reconnaissance missions offer great contributions to a better understanding of the origins of our Solar System, planetary defense from hazardous objects and potential economic return from its rare resource extraction. The employment of solar sail propelled CubeSats enable a greater number of missions and target asteroids, given it is a cheaper propulsion method. However, modeling the dynamics of solar sails orbiting irregular gravitational fields is a challenge. And since asteroids have a wide variety of shapes and sizes, it is important to make use of a robust optimization method that allows fast adjustments to this dynamics. This study examines the use of different solar sails and asteroid sizes for station-keeping a spacecraft orbiting an irregularly shaped asteroid. To achieve this goal, a rational agent was developed to act as an attitude guidance system. It calculates the necessary sail orientation to keep the spacecraft near its initial orbit. The agent uses a direct optimization strategy based on a decision tree algorithm in order to account for a solar sail with non-ideal reflective properties and the non-uniform gravitational field of the asteroid. Additionally, a constraint for the attitude change frequency is imposed to account for the difficult maneuverability of solar sails. By testing multiple sail sizes, this study evaluates how long the agent may maintain the same sail orientation while still being successful at its station keeping mission. This analysis is done for different asteroid sizes and a relation between different sail and asteroid sizes is established. Results reveal a relation between sail and asteroid sizes and the attitude change frequency. Smaller sails can sustain station-keeping with fewer attitude changes. However, they are not able to achieve this goal on larger asteroids. For sufficiently small asteroids, sails smaller than those used on recent missions (e.g.: LightSail-2, NEA-Scout, ACS3) may be more appropriate for station-keeping. These findings suggest that current solar sail technology is already capable of enabling successful asteroid reconnaissance missions.
\end{abstract}

\keywords{Orbital mechanics \and Solar sailing \and Asteroid reconnaissance \and Rational agent}

\section{Introduction}
\label{sec:intro}

Asteroids are a means of better understanding the origins of our Solar System. Their composition and orbital properties offer insights on the formation of Earth and all the other planets and moons~\citep{kareta:2025}. Additionally, a detailed knowledge of their dynamics provides a better chance of defense from possible collision scenarios~\citep{chabot:2024}. In the future, they could also be mined for their rare metals, serving as a source of economic growth and greater technological prosperity~\citep{kendal:2025}.

In this context, space agencies have a growing interest in asteroid reconnaissance missions. Some examples are the OSIRIS-REx/APEX~\citep{burns:2025,dellagiustina:2023} or Lucy~\citep{olkin:2024} missions from the National Aeronautics and Space Administration (NASA), Hera~\citep{michel:2022} from the European Space Agency (ESA) and Hayabusa missions~\citep{yoshikawa:2021,watanabe:2024} from the Japanese State Space Agency (JAXA). When considering the vast number of asteroids in our Solar System, cheaper means of propulsion would mean that a larger quantity of missions could be developed, resulting in a greater number of information intake from these bodies. This is why it is interesting to investigate the use of solar sails spacecrafts to perform asteroid reconnaissance. Solar sails have been tested as a form of propulsion to CubeSats in Leo Earth Orbit~\citep{vulpetti-2015,spencer-2021,wilkie:2023} and using them to investigate Near Earth Asteroids, as intended from the failed NEA-Scout mission~\citep{johnson-2022}, is a possible next step to this technology.

This study proposes the use of a solar sail to station keep a spacecraft orbiting an irregularly shaped asteroid. Different sized asteroids and solar sails are considered. Given the large variety of shapes and sizes asteroids can have, added to the complexity of orbital dynamics with a non-homogeneous gravitational field and the solar radiation pressure acceleration from a non-ideally reflecting solar sail, it is beneficial to implement a direct optimization method, known for cheaper and faster implementation~\citep{shirazi:2018}. A rational agent is used as an attitude guidance system~\citep{meireles:2026-ABCM} to determine the necessary sail orientation throughout the mission, with the objective of maintaining the spacecraft near its initial orbital conditions for the total duration of the mission. \added{Finally, this study analyzes the relation between the asteroid size, the sail size and the interval between attitude changes in order to successfully station keep the spacecraft in its intended orbit.}

\added{The agent is the product of a newly explored method for the design of solar sail missions. The use of artificial intelligence techniques allows a fast formulation of complex system dynamics of multiple mission scenarios. It also sustains a formulation with multiple objectives and is able to consider the attitude changes of the sail as a cost, which is a novelty approach for solar sail mission design. The agent has been recently applied in the use of Low Earth Orbits, considering the atmospheric drag, and Low Lunar Orbits, considering up to $20$~orders of spherical harmonics expansion of the gravitational potential of the Moon. It is a tool under constant development and it is crucial to test it under different mission scenarios, such as station keeping over irregularly shaped asteroids. Given that the main focus of this study was the analysis of the asteroid and sail size relation and the agent still presents a few unknown intricacies, the system formulation of this study was simplified (i.e.: asteroid with ellipsoidal shape). However, from the gained experience of this study and an upgradable algorithm of the agent, future studies will be able to explore more elaborate dynamics (e.g.: shapes which are even more irregular or binary asteroid systems).}

\section{System Dynamics}
\label{sec:theory}

\added{This section is dedicated to presenting the models considered for the gravitational and solar radiation pressure accelerations (Section~\ref{sec:theory/sysMod}), the initial conditions explored throughout this study (Section~\ref{sec:theory/orbCond}) and the natural trajectory these initial conditions offer for a spacecraft without the propulsion from a solar sail (Section~\ref{sec:theory/noSail}).}

\subsection{System models}
\label{sec:theory/sysMod}

This study considers a spacecraft subject to the acceleration from a non-homogeneous gravitational field of an ellipsoidal asteroid $(\mathbf{a}_{g,ast\rightarrow sc})$ and the gravitational acceleration from the Sun $(\mathbf{a}_{g,\text{\Sun}\rightarrow sc})$. Additionally, the spacecraft has as its source of propulsion the acceleration from the solar radiation pressure $(\mathbf{a}_{SRP})$. In this manner, the total acceleration of the spacecraft is:

\begin{equation}
\mathbf{a}_{sc} = \mathbf{a}_{g,ast\rightarrow sc} + \mathbf{a}_{g,\text{\Sun}\rightarrow sc} + \mathbf{a}_{SRP}
\label{eq:accTotal}
\end{equation}

The system is numerically integrated as a 3-Body Problem, in a Heliocentric Inertial Frame (HIF). The HIF is centered in the position of the Sun and its XY-plane defines the ecliptic. The numerical integration of the system is performed with the coordinates of all the bodies in the HIF. All the gravitational accelerations, with the exception of the acceleration of the spacecraft from the asteroid, are considered from a point mass model:

\begin{equation}
\mathbf{a}_{g,b_i\rightarrow b_j} = -\frac{\mu_{b_i}}{r^3_{b_i\rightarrow b_j}}\mathbf{r}_{b_i\rightarrow b_j}
\label{eq:accSun}
\end{equation}
where $\mathbf{a}_{g,b_i\rightarrow b_j}$ is the gravitational acceleration \added{that} a body $b_j$ suffers from a body $b_i$, $\mu_{b_i}$ is the gravitational parameter of body $b_i$ and $\mathbf{r}_{b_i\rightarrow b_j}$ is the distance vector between the two bodies.

In this manner, the equations of motion of the system are:
\begin{equation}
\left\{ \begin{aligned}
\ddot{x}_\text{\Sun} &= \mathbf{a}_{g,ast \rightarrow \text{\Sun}} + \mathbf{a}_{g,sc \rightarrow \text{\Sun}} \\
\ddot{x}_{ast} &= \mathbf{a}_{g,\text{\Sun} \rightarrow ast} + \mathbf{a}_{g,sc \rightarrow ast} \\
\ddot{x}_{sc}  &= \mathbf{a}_{g,\text{\Sun} \rightarrow sc} + \mathbf{a}_{g,ast \rightarrow sc} + \mathbf{a}_{SRP} \\
\end{aligned}\right.
\label{eq:sisMotion}
\end{equation}

\begin{figure}[!ht]
    \centering
    \includegraphics[width=0.8\columnwidth]{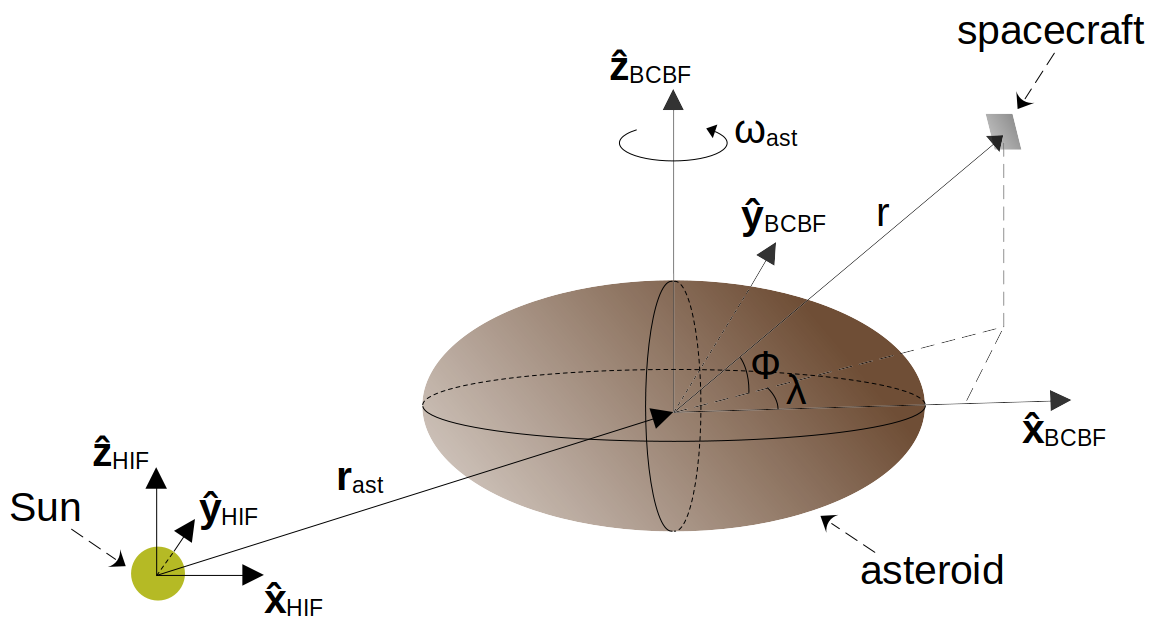}
    \caption{Body-Centered Body-Fixed Frame (BCBF).}
    \label{fig:BCBF}
\end{figure}

The acceleration from a non-homogeneous gravitational field was calculated from a spherical harmonics expansion:
\begin{equation}\label{eq:accMB}
\mathbf{a}_{g,ast\rightarrow sc} = -\frac{\mu}{r^2} \left\lbrace 1+ \sum\limits_{n=1}^{\infty} \sum\limits_{m=0}^{n} \left( \frac{R_{ref}}{r} \right) ^n {P}_{nm} \left( \sin(\phi) \right) \left[ {C}_{nm} \cos(m\lambda) + {S}_{nm} \sin(m\lambda) \right] \right\rbrace \hat{\mathbf{r}}
\end{equation}
where $\mu$ and $R_{ref}$ are, respectively, the gravitational parameter and reference radius of the central body. The pair $(n,m)$ indicate the degree and order of the associated Legendre functions $(P_{nm})$ and the spherical harmonics coefficients $(C_{nm},S_{nm})$. Finally, $(r,\lambda,\phi)$ are polar coordinates of the spacecraft defined in a frame fixed on the central body \added{(Body-Centered Body-Fixed Frame - BCBF)}. The axes of this reference frame are defined from the three main semi-axes of the asteroid, where the z-axis is also the rotational axis of the body. The spacecraft coordinates in this frame are only used to calculate the spherical harmonics expansion, whereas the actual integration of the problem is done in the HIF. \added{An illustration of the BCBF frame is presented in Figure~\ref{fig:BCBF}.}

This approach allows a fast adaptation of the dynamics to all the possible different shapes and sizes of asteroids. This study considered an ellipsoid shaped asteroid, as a means of obtaining more general results. However, for different shaped asteroids, the agent only requires a change of the coefficient values $(C_{nm},S_{nm})$ to perform its calculations.

An ellipsoid may be described from its three main semi-axes: $\mathbf{R}_{ast} = [\,R_{1}\,,\,R_{2}\,,\,R_{3}]$. The oblateness is represented by the $C_{20}$ coefficient and the ellipticity by the $C_{22}$ coefficient~\citep{balmino:1994}:
\begin{equation}
C_{20} = (2R_{3}^2 - R_{1}^2 - R_{2}^2)/10R_{ref}^2
\label{eq:C20}
\end{equation} 
\begin{equation}
C_{22} = (R_{1}^2 - R_{2}^2)/20R_{ref}^2
\label{eq:C22}
\end{equation}
where $R_{ref} = (R_{1}R_{2}R_{3})^{1/3}$. The other spherical coefficients are null.

\begin{figure}[!ht]
    \centering
    \includegraphics[width=0.8\columnwidth]{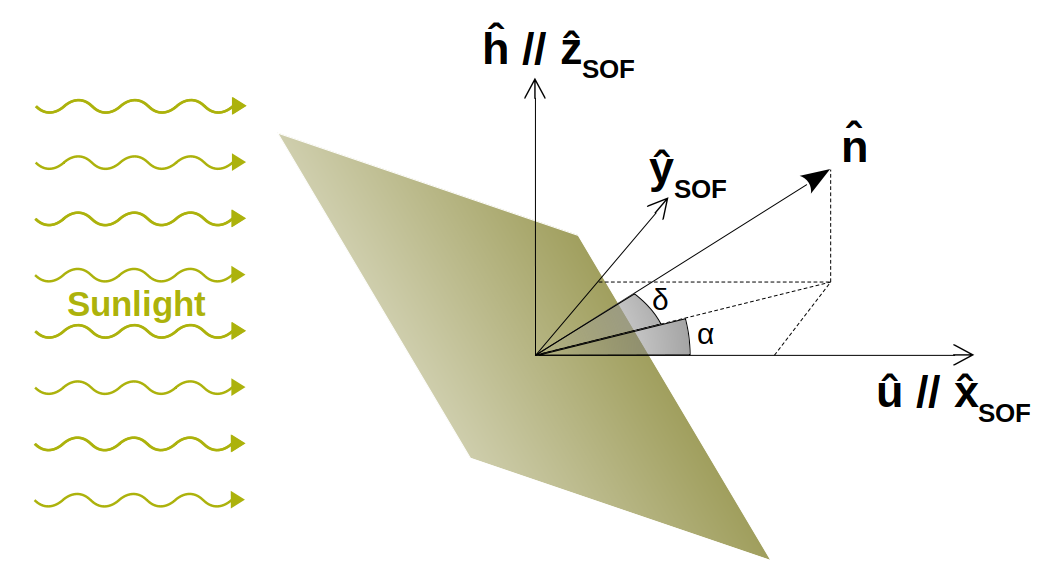}
    \caption{Spacecraft Oriented Frame (SOF).}
    \label{fig:SOF}
\end{figure}

Additionally, the solar radiation pressure acceleration resulting from a solar sail is calculated from:
\begin{equation}
\mathbf{a}_{SRP} = \ \mathbf{\Xi} \left( \frac{\mu_\text{\Sun}}{r_{\text{\Sun}\rightarrow sc}^2} \right) \boldsymbol{\ell}
\label{eq:accSRP}
\end{equation}
where $\mu_\text{\Sun} = 1.32712440018\times10^{20}\,\mathrm{m^3/s^2}$ is the gravitational parameter of the Sun and $\boldsymbol{\ell}$ is the lightness vector~\citep{vulpetti:2015}. Since $\boldsymbol{\ell}$ is defined in a Spacecraft Oriented Frame (SOF), depicted in Figure~\ref{fig:SOF}, it is necessary to use a direction cosine matrix $\mathbf{\Xi}$ to calculate $\mathbf{a}_{SRP}$ in the HIF.
This transformation matrix is defined as:
\begin{equation}
\mathbf{\Xi} = \begin{bmatrix}
\hat{\mathbf{x}}_{SOF} \\
\hat{\mathbf{y}}_{SOF} \\
\hat{\mathbf{z}}_{SOF} \\
\end{bmatrix}^T
\label{eq:rotMat}
\end{equation}
where $\left( \hat{\mathbf{x}}_{SOF},\hat{\mathbf{y}}_{SOF},\hat{\mathbf{z}}_{SOF} \right)$ are the Spacecraft Oriented Frame (SOF) main axis, which are defined from the radial heliocentric direction $\left( \hat{\mathbf{r}}_{\text{\Sun}\rightarrow sc} \right)$ and the direction of the spacecraft heliocentric angular momentum $\left( \hat{\mathbf{h}}_{\text{\Sun}\rightarrow sc} \right)$ in such a manner: $\hat{\mathbf{x}}_{SOF} = \hat{\mathbf{r}}_{\text{\Sun}\rightarrow sc}$, $\hat{\mathbf{y}}_{SOF} = \hat{\mathbf{h}}_{\text{\Sun}\rightarrow sc}\!\times\!\hat{\mathbf{r}}_{\text{\Sun}\rightarrow sc}$ and $\hat{\mathbf{z}}_{SOF} = \hat{\mathbf{h}}_{\text{\Sun}\rightarrow sc}$. 

On the other hand, the lightness vector $(\boldsymbol{\ell})$ is a function of the sail attitude $(\hat{\mathbf{n}})$ and its optical properties:
\begin{equation}
\boldsymbol{\ell} = \left( \frac{1}{2} \frac{\sigma_c}{\sigma} \right) n_x [( 2r_\mathrm{spec} n_x+\chi_f r_\mathrm{diff} +
\kappa a_\mathrm{b} ) \hat{\mathbf{n}} + ( a_\mathrm{b} + r_\mathrm{diff} ) \hat{\mathbf{u}} ]
\label{eq:light}
\end{equation}
where $\hat{\mathbf{u}}$ is the unit vector oriented in the direction of the sunlight and $\sigma_c = 1.5368\,\mathrm{g/m^2}$ is a constant named critical sail loading~\citep{mcinnes-2004}. Additionally, $\sigma = m_{sc}/S_{sail}$ is the sail loading, which is the ratio of the total spacecraft mass $(m_{sc})$ to the sail surface area $(S_{sail})$. In turn, the sail area $S_{sail} = 2\,l_{boom}^2$ is calculated from the length $(l_{boom})$ of the four booms which support the opened sail. The term $n_x$ is the component of $\hat{\mathbf{n}}$ in the $\hat{\mathbf{x}}_{SOF}$ direction. Finally, the optical properties were taken from~\citet{heaton:2015}, where $r_{spec} = 0.8272$, $r_{diff} = 0.0528$, $a_b = 0.12$ and $\chi_f = 0.79$ are, respectively, the specular reflectance, diffuse reflectance, absorptance and non-Lambertian coefficients of the reflective side of the sail and $\kappa = -0.4383$ is an emission/diffusion net thrust factor.

Finally, it is possible to express the sail attitude $(\hat{\mathbf{n}})$ from two angles: the azimuth $(\alpha)$ and elevation $(\delta)$, which are also shown in Figure~\ref{fig:SOF}. The azimuth is the angle between the projection of $\hat{\mathbf{n}}$ on the xy-plane of the SOF and $\hat{\mathbf{x}}_{SOF}$. The elevation is the angle between $\hat{\mathbf{n}}$ and the xy-plane of the SOF.
\begin{equation}
\hat{\mathbf{n}} \equiv 
\begin{bmatrix}
\cos(\alpha) \cos(\delta) \\
\sin(\alpha) \cos(\delta) \\
\sin(\delta) \\
\end{bmatrix}
\label{eq:sailNormal}
\end{equation}

\subsection{Initial conditions}
\label{sec:theory/orbCond}

The spacecraft is initialized at a circular polar orbit with a dusk-dawn configuration. Although the problem is not Keplerian, the initial cartesian coordinates are calculated from a direct transformation of the Keplerian elements of an osculating orbit around a body with the gravitational parameter of the asteroid $(\mu_{ast})$. This configuration is chosen to maintain the sail exposed to the sunlight at all moments of a revolution. In this manner, the initial orbit eccentricity, inclination and longitude of ascending node are, respectively, $KE_e(t_0) = 0.0$, $KE_i(t_0) = 90^\circ$, $KE_\Omega(t_0) = 270^\circ$.

A set of two conditions are modified throughout the study:
\begin{enumerate}
    \item The asteroid dimensions:\\
The shape of the asteroid is defined from the following ellipsoidal semi-axis: $\mathbf{R}_{ast} = [\,500.0\,,\,250.0\,,\,250.0\,]\,\mathrm{m}$. The semi-axis ($\mathbf{R}_{ast}$) is then multiplied by ten to the power of a dimension factor $\left( \times10^{f_{dim}} \right)$, where $f_{dim} = (0,1,2,3)$. This defines four asteroids with the same shape but different dimensions.
    \item The sail size:\\
Four sail boom lengths are considered: $l_{boom} = [\,1.0\,,\,3.0\,,\,5.0\,,\,7.0\,]\,\mathrm{m}$. By considering a fixed total spacecraft mass of $m_{sc} = 16.0\,\mathrm{kg}$, this results in four different sail loading sizes: $\sigma = [\,8.0\times10^3\,,\,888.89\,,\,320.0\,,\,163.27\,]\,\mathrm{g/m^2}$. Both $m_{sc}$ and $l_{boom} = 7.0\,\mathrm{m}$ values were taken from the ACS3 mission from NASA~\citep{wilkie:2023}.
\end{enumerate}

Additionally, the gravitational parameters of the asteroids are determined from their respective ellipsoidal volumes: $\mu = G \rho_{ast} \left( \frac{4}{3}\pi R_1 R_2 R_3 \right)$, where $G = 6.67430\times10^{-11}\,\mathrm{m^3/kg/s^2}$ is the Universal Gravitational Constant and $\rho_{ast} = 2.0\times10^3\,\mathrm{kg/m^3}$ is the mean density of the asteroids, taken from a range of common values for this property~\citep{carry:2012}.

The initial semi-major axis of the spacecraft ($KE_{a}(t_0)$) for each of the four asteroids was calculated as a function of their respective reference radius: $KE_{a}(t_0) = 2.0 \times R_{ref}$, where $R_{ref} = (314.98 \times 10^{f_{dim}})\,\mathrm{m}$. All of these orbits result in an initial orbital period of $T_0 \approx 6h\,36min$. The total mission time ($t_f$) was defined from the initial orbital period: \added{$t_f = 30\times T_0$}. The motive for this value is better explained in Section~\ref{sec:theory/noSail}.

In turn, the asteroid is initialized at the perihelion of an heliocentric orbit with zero inclination, semi-major axis equal to $2.0\,\mathrm{au}$ and an eccentricity of $0.5$. This results in an initial distance of $1.0\,\mathrm{au}$, that slightly increases over the course of the simulation. These orbital parameters were defined as a generic Near-Earth Asteroid orbit. It is important to note that different initial positions or orbits have the potential of altering the results, because the $\mathbf{a}_{SRP}$ is a function of the distance of the sail to the Sun. However, this affects the results in a similar way the sail size does, by indirectly varying the magnitude of $\mathbf{a}_{SRP}$. In order to limit the number of different conditions studied, different orbital parameters of the asteroid were not analyzed, but the relation between these variations and different sail sizes can be the focus of future studies.

The asteroid has a rotational period of $12$~hours, which is slightly smaller than $2.0 \times T_0$. Even though it was chosen as an arbitrary value, this relation with $T_0$ distinguishes it from the extremes of either a ``too fast'' or a ``too slow'' rotational period. Different rotational periods result in different natural trajectories (without a solar sail), which affect the resulting attitudes calculated by the agent when using a solar sail. However, the main focus of this study are not the attitude values by themselves, but actually the capacity of the agent to maintain the orbit with the specified solar sail specifications. In this manner, it is not interesting to explore different rotational periods in this study. The same assumptions can be made regarding the angle between the asteroid rotational axis and the orbital plane of the spacecraft. The asteroid has an axial tilt of $0^\circ$ and different values were not explored in this study.

In short, the initial cartesian coordinates of the bodies are taken from the osculating Keplerian elements indicated in Table~\ref{tab:initCond}. The Sun is initialized in the center of the HIF, with position and velocity equal to zero.

\begin{table*}
\centering
\caption{Initial orbital conditions $KE(t_0)$}
\begin{tabular}{|c|c|c|c|c|c|c|}
\hline
$\text{Body}_{\rightarrow \text{``Central Body''}}$ & $KE_{a}$ & $KE_{e}$ & $KE_{i}$ & $KE_{\Omega}$ & $KE_{\omega}$ & $KE_{f}$ \\
\hline
$ast_{\rightarrow \text{\Sun}}$ & $2.0\,\mathrm{au}$ & $0.5$ & $0^\circ$ & $0^\circ$ & $0^\circ$ & $0^\circ$\\
\hline
$sc_{\rightarrow ast}$ & $2.0 \times R_{ref}$ & $0.0$ & $90^\circ$ & $270^\circ$ & $0^\circ$ & $0^\circ$ \\
\hline
\end{tabular}
\label{tab:initCond}
\end{table*}

It is important to reinforce that four different initial orbits for the spacecraft are considered, with four different radius (one for each asteroid size, determined from $R_{ref}$), but all the other parameters are equal. In spite of the different radius, given that $KE_{a}(t_0) \propto R_{ref}$ and $\mu_{ast} \propto R_{ref}^{1/3}$, the ratio $\left( \frac{KE_{a}(t_0)}{\mu_{ast}^{1/3}} \right)$ is constant, which offers the same dynamical conditions. However, different conditions would arise for initial altitudes different than $2.0 \times KE_a(t_0)$, such as $2.5 \times KE_a(t_0)$ or $3.0 \times KE_a(t_0)$. The results obtained by the agent would be different but, once again, this analysis was discarded because it would slightly shift the focus of this study.

\subsection{Natural trajectory}
\label{sec:theory/noSail}

A trajectory without a sail, for the same initial conditions of Table~\ref{tab:initCond}, was calculated as a reference. As already explained in Section~\ref{sec:theory/orbCond}, given that the initial altitude is determined as a function of $R_{ref}$, even though different values of $f_{dim}$ result in different $\mu_{ast}$, the trajectories are all the same if all the other variables are maintained. In this case, it is sufficient to analyze the trajectory of $f_{dim} = 3$, which is presented in this section.

\begin{figure}[!ht]
    \centering
    \includegraphics[width=0.6\columnwidth]{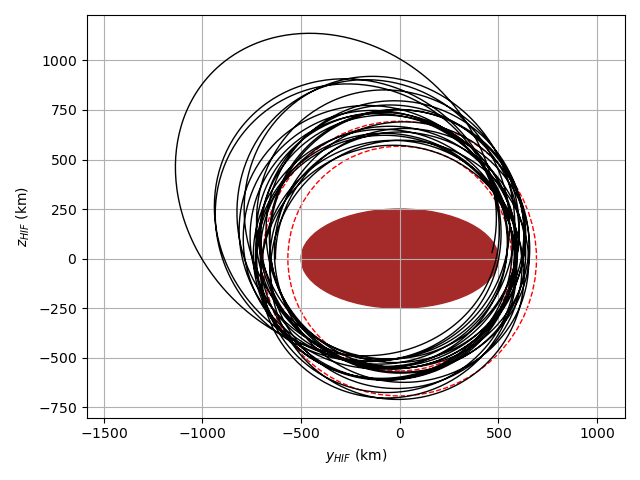}
    \caption{Trajectory without sail.}
    \label{fig:exOff_traj}
\end{figure}

The simulation was interrupted at $t = t_{coll}$ when a collision with the asteroid occurred, such that $\frac{r_{x}(t)^2}{R_1^2} + \frac{r_{y}(t)^2}{R_2^2} + \frac{r_{z}(t)^2}{R_3^2}< 1.0$. Figure~\ref{fig:exOff_traj} illustrates this trajectory. It also shows the upper and lower distance limits in red dashed circles.

The ellipse in the center of Figure~\ref{fig:exOff_traj} represents a stationary image of the central body. However, it is an elongated body that rotates around its z-axis. Therefore, an intersection of the trajectory (black line) over the colored area of the ellipse does not necessarily characterize a collision.

\begin{figure}[!ht]
    \centering
    \includegraphics[width=0.6\columnwidth]{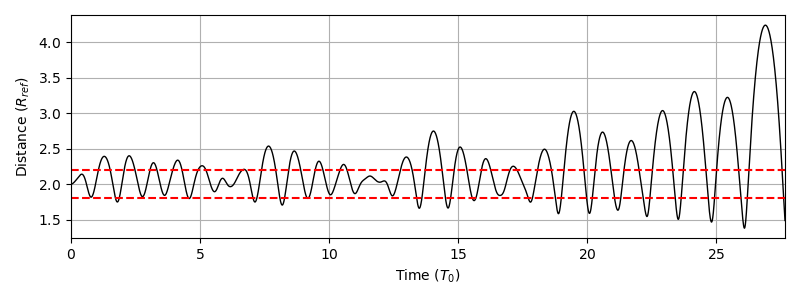}
    \caption{Spacecraft distance to the center of the asteroid, as a function of time, from the trajectory without sail.}
    \label{fig:exOff_dist}
\end{figure}

Figure~\ref{fig:exOff_dist} shows the spacecraft distance to the center of the asteroid as a function of time, while also displaying the upper and lower distance limits in red dashed lines. It can be seen that the first time the final goal is disrespected is at $t = 1.07 \times T_0$. This trajectory has a maximum distance of $max(r) = 4.24 \times R_{ref}$ and the collision occurs at $t_{coll} = 27.67 \times T_0$. Therefore, a mission of \added{$t_f = 30 \times T_0$} outlasts a scenario where no solar sail is employed.
\section{Station keeping algorithm}
\label{sec:method}

\added{This section is dedicated to presenting the algorithm used to calculate the trajectories to station keep a solar sail spacecraft around an irregularly shaped asteroid. The algorithm is based in artificial intelligence techniques and Section~\ref{sec:method/agent} discusses the overall structure of the rational agent developed to perform the intended calculations. The search technique employed by the agent is explained in more detail in Section~\ref{sec:method/searchTech} and its goals and heuristic formulation in Section~\ref{sec:method/heuristic}.}

\subsection{Rational Agent}
\label{sec:method/agent}

The modeling of the rational agent and its search process is inspired by~\citet{russell:2016} and is described throughout this section. It is a multipurpose search tool, previously developed by \citet{meireles:2026-ABCM}, that acts as an attitude guidance system for a solar sail propelled spacecraft. Its versatile structure allows the goal state to be adjusted depending on the mission specifications. \added{The agent is under constant development and this study was an opportunity to explore its application in station keeping solar sails around irregularly shaped asteroids.}

\begin{figure}[!h]
    \centering
    \includegraphics[width=0.6\columnwidth]{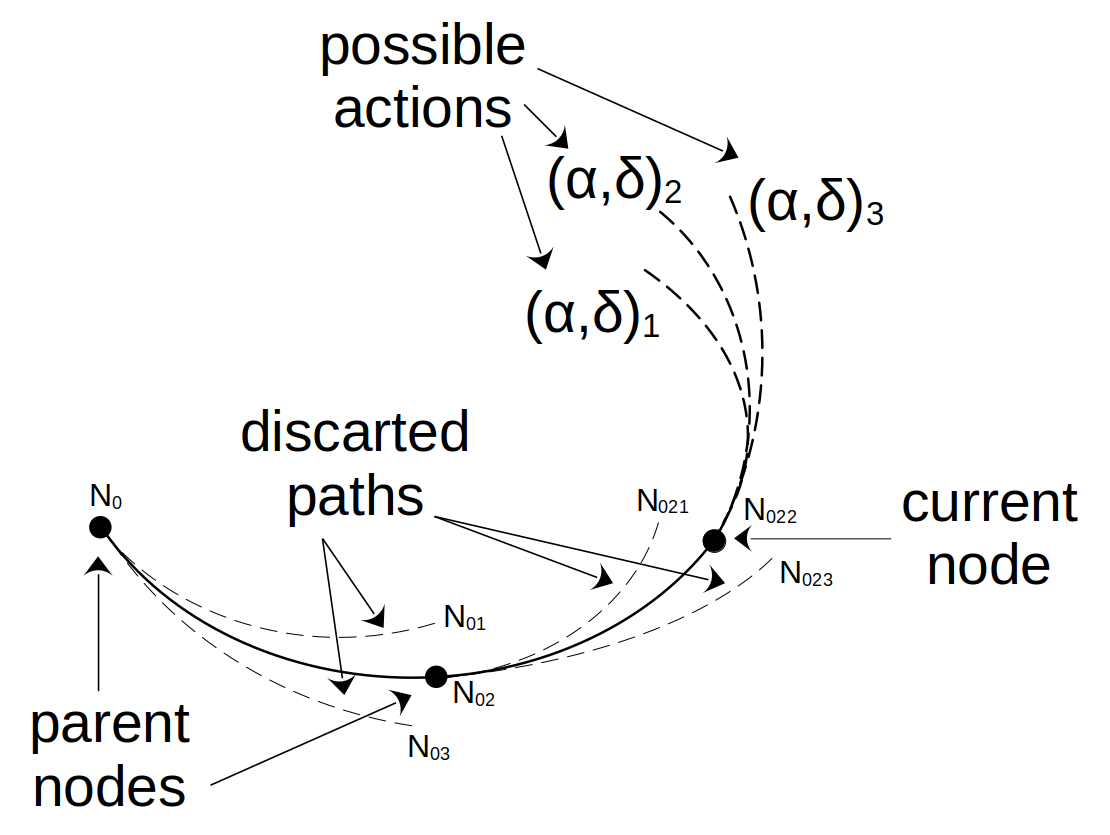}
    \caption{Agent search diagram.}
    \label{fig:agentDiag}
\end{figure}

In short, the agent is able to predict the consequences of its actions by numerically integrating Equation~\ref{eq:accTotal}, where an action is considered as a specific sail attitude defined from $(\alpha,\delta)$. Different attitudes lead the spacecraft to different coordinates, with its associated osculating orbital states. Each of these different coordinates is referred to as a decision node $(N)$. A single node has its current state integrated for a $\Delta t(N)$ time span, for all the different actions considered, generating multiple children nodes with different coordinates. This means that, at every $\Delta t(N)$ time interval, the agent decides on a new sail attitude. By the same process, each of these children nodes are able to generate their own successors, all resulting in new coordinates. This process is illustrated in Figure~\ref{fig:agentDiag}. In the diagram, there is a root node $N_{0}$ with three children nodes: $N_{01}$, $N_{02}$ and $N_{03}$. In each of them, the spacecraft has different coordinates generated from different sail attitudes. The first node and the latter were discarded and node $N_{02}$ generated its own children nodes: $N_{021}$, $N_{022}$ and $N_{023}$. The paths the spacecraft goes through are illustrated by the gray dashed lines for discarded nodes and solid black line for the selected nodes. Finally, node $N_{022}$ is the current node under evaluation, where three different attitudes lead to three new possible coordinates which will be evaluated and compared.

Additionally, every node $N$ is assigned a heuristic value ($h(N)$), which is an estimate of the cost to reach the final goal. These values are defined in Section~\ref{sec:method/heuristic}. In this manner, nodes can be ordered as most-least promising to achieve the final goal and given priority in the generation of future nodes. This is repeated until a node fulfills the final goal conditions. The sequence of actions/sail attitudes that lead to the generation of this node is returned as the answer to this problem. 

Every node is an independent object, with all the information needed to perform its own integration. Its attributes are the current orbital state, the integration time, the past actions list which lead to its creation and its heuristic value. The code was implemented with the use of the Pathos Framework~\citep{mckerns:2012}, in Python programming language, to parallelize integrations of different nodes. The search algorithm operates at an upper layer and is better explained in Section~\ref{sec:method/searchTech}. It creates, sorts, manipulates and discards nodes in different manners, depending on the search technique applied.

\subsection{Search technique}
\label{sec:method/searchTech}

In this study, the node integration time ($\Delta t(N)$) was determined as a function of the initial orbital period. A range of values was considered: $\Delta t(N) = [\,\,0.5\,,\,1.0\,,\,2.0\,,\,4.0\,,\,8.0\,] \times T_0$. The objective was finding a solution that respects the goal at all times with the greatest $\Delta t(N)$ value, which means fewer attitude changes throughout the mission.

Each node considers an action grid step of $\Delta \alpha = \Delta \delta = 15^\circ$ to determine their successors. Since $\alpha$ and $\delta$ belong to the interval $[-90^\circ,90^\circ]$ and every combination of $\alpha$ or $\delta$ equal to $-90^\circ$ or $90^\circ$ results in a null $\mathbf{a}_{SRP}$, all of these combinations except one can be excluded from the action list of a node. Consequently, a $15^\circ$ grid step results in $(11^2 + 1) = 122$~children nodes.

In respect to the size of the search space, the worst case scenario would be the smallest node integration time considered $(\Delta t(N) = 0.5 \times T_0)$. For an integration time of \added{$t_f = 30 \times T_0$}, this results in a search tree of $60$ levels. Given that each node generates $122$ successors, this means a total of $122^{60}$ nodes.

In order to deal with the vast search space of this problem, it is necessary to apply a search technique that offers a balance between memory usage and optimality of the solution. In this case, a Beam Search~\citep{russell:2016} was used to calculate the solution of each scenario.

In a beam search, the agent selects a list of the best $W$ successor nodes, where $W$ is the beam width and the term ``best'' refers to the nodes with lowest heuristic $(h(N))$. It then calculates all the $(122 \times W)$ successors and, once again, selects the best $W$ nodes within this pool. It repeats this process until $(t = t_f)$. In this manner, it is a greedy algorithm. This study considered a beam width of $W = 100$, which means selecting approximately $1\%$ of the best nodes from all the successors created at each level.

\subsection{Heuristic}
\label{sec:method/heuristic}

The proposed final goal of this study is finding a trajectory, with a total duration $t_f$, where the distance to the center of the asteroid ($r(t) = |\mathbf{r}(t)|$) does not vary more than $10\%$ of the initial distance ($r(t_0)$) while also maintaining a polar dusk-dawn configuration:
\begin{equation}
goal \rightarrow \left\{ \begin{aligned}
&\frac{max(|\,r(t) - r(t_0)\,|)}{r(t_0)} < 10\% \\
&KE_{i}(t) = 90^\circ \pm 1^\circ \\
&KE_{\Omega}(t) = (270^\circ + KE_{f}^{\text{\Sun}\rightarrow ast}(t)) \pm 1^\circ
\end{aligned}\right.
\label{eq:goals}
\end{equation}
where $KE_{f}^{\text{\Sun}\rightarrow ast}(t)$ is the heliocentric true anomaly of the asteroid.

In order to respect the imposed restrictions, the heuristic function is:
\begin{equation}
h(N) = w_{r}\frac{|\,r(N) - goal_{r}|}{tol_{r}} + w_{i}\frac{|KE_{i}(N) - goal_{i}|}{tol_{i}} + w_{\Omega}\frac{|KE_{\Omega}(N) - goal_{\Omega}|}{tol_{\Omega}}
\end{equation}
where $\mathbf{goal} = [\,goal_{r}\,,\,goal_{i}\,,\,goal_{\Omega}\,] = [\,r(t_0)\,,\,90^\circ\,,\,(270^\circ+KE_{f}^{\text{\Sun}\rightarrow ast}(t))\,]$ are the goal values for the distance, inclination and longitude of ascending node. The last one must be updated according to the heliocentric position of the asteroid. Additionally, $\mathbf{tol} = [\,tol_{r}\,,\,tol_{i}\,,\,tol_{\Omega}\,] = [\,10\%\cdot r(t_0)\,,\,1^\circ\,,\,1^\circ\,]$ are the tolerances used by the agent. Finally, weighting factors of $\mathbf{w} = [\,w_{r}\,,\,w_{i}\,,\,w_{\Omega}\,] = \left[\,\frac{1}{3}\,,\,\frac{1}{3}\,,\,\frac{1}{3}\,\right]$ were applied to guarantee that $h(N) < 1$ if all the states lie within the admitted tolerances. 

Depending on the objectives of the study, or dynamical limitations of the problem, the weights might be adjusted to prioritize the agent focus on one term over the other. \added{Overall, the choice of $\mathbf{tol}$ and $\mathbf{w}$ are made from a set of test simulations, where the quality of the solutions are compared for different $\mathbf{tol}$ and $\mathbf{w}$ values. For example, if a goal is less easily reached than another, a greater weight is addressed to it. Considering a station keeping objective, where all values are already within range at the start of the simulation, there is no reason to prioritize one goal over the other with different weighting factors. On the other hand, the solar sail size creates a limitation for the values of $\mathbf{tol}$. If any of their terms are too small, some of the goals are easily disrespected and the $h(N)$ of the descendants become too depreciated, making it harder for the agent to evaluate between equally poor choices. In this case, the agent starts to prioritize some goals and ignore others. However, these are extreme cases for very small tolerances. Results from this study did not vary significantly if bigger tolerances were chosen and the values considered were deemed reasonable to corroborate a successful station keeping.}

The quality of orbit maintenance ($\eta$) is measured as the sum of every time interval where the goals from Equation~\ref{eq:goals} are respected. The sum of these intervals is then divided by the total integration time (\added{$30\times T_0$}):
\begin{equation}
\eta = \frac{1}{30T_0}\sum_{i=1}^n (t_{f,i}-t_{0,i}) \ ,\quad \mathrm{if\ goal\rightarrow\ True\ for\ (t_{0,i} < t < t_{f,i})}
\end{equation}
where $n$ is the amount of time intervals that respect such conditions.


\section{Solar sailing with a rational agent}
\label{sec:res/wSail}

The agent employed the different sized solar sails over the different asteroids for a mission duration of \added{$t_f = 30 \times T_0$}, which is slightly longer than the collision time without a solar sail $(t_{coll} = 27.67 \times T_0)$. If the spacecraft survives the mission duration, it can already be considered a successful use of a solar sail.

\subsection{Distance maintenance}
\label{sec:res/dist}

An initial analysis focuses on how effective the agent was in maintaining the distance within $10\%$ of the initial distance $(r_0)$. The largest orbit maintenance ($\eta$), only for $goal_r$, and their respective attitude change intervals $(\Delta t(N))$ are displayed in Table~\ref{tab:results/cond1}.

\begin{table*}
\centering
\caption{Decision time $\Delta t(N)$, measured in $T_0$, with the largest $\eta$, in parenthesis, for $goal_r \rightarrow$ True}
\begin{tabular}{|c|*{4}{c|}}
\hline
\diagbox[]{$f_{dim}$}{$l_{boom}\,\mathrm(m)$} & $1.0$ & $3.0$ & $5.0$ & $7.0$ \\ 
\hline $0$
& $2.0\ (67.62\%)$ & $1.0\ (79.15\%)$ & $1.0\ (89.30\%)$ & $1.0\ (71.17\%)$ \\
\hline $1$
& $8.0\ (68.68\%)$ & $2.0\ (71.97\%)$ & $1.0\ (93.29\%)$ & $1.0\ (95.29\%)$ \\
\hline $2$
& N/A & $4.0\ (67.51\%)$ & $1.0\ (67.61\%)$ & $2.0\ (66.34\%)$  \\
\hline $3$
& N/A & N/A & $8.0\ (52.52\%)$ & $8.0\ (66.53\%)$  \\
\hline
\end{tabular}
\label{tab:results/cond1}
\end{table*}

Initially, a threshold can be identified from the ``N/A'' results: there are scenarios where the sail is too small for the size of the asteroid. The sail does not offer sufficient $\mathbf{a}_{SRP}$ to maintain the orbit inside the acceptable region, or even make sufficient deviations from the reference trajectory (without sail). As a consequence, all the $\Delta t(N)$ evaluated resulted in collisions.

\begin{figure}[!ht]
    \centering
    \includegraphics[width=0.6\columnwidth]{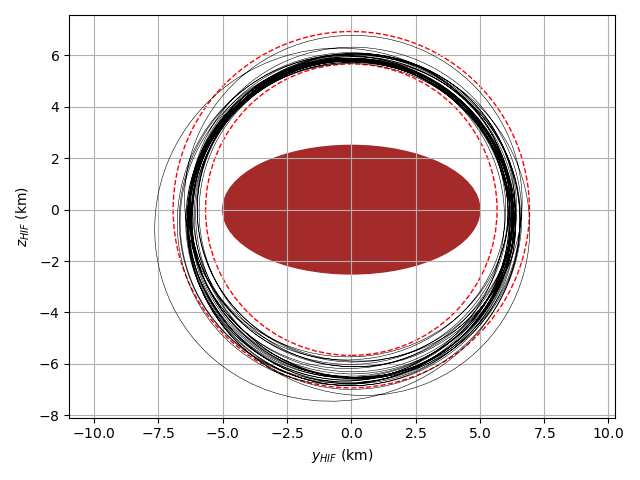}
    \caption{Trajectory for $f_{dim} = 1$, $l_{boom} = 7.0\,\mathrm{m}$ and $\Delta t(N) = 1.0 \times T_0$.}
    \label{fig:ex171_traj}
\end{figure}

Overall, in the successful cases, the agent was capable of maintaining the distance within the tolerance range for at least $66\%$ of the time, with an outlier of $52\%$ at an extreme case $(f_{dim}  = 3$ and $l_{boom} = 5.0\,\mathrm{m})$. The best case resulted in a maintenance of over $95\%$ $(f_{dim}  = 1$ and $l_{boom} = 7.0\,\mathrm{m})$. Its trajectory is illustrated in Figure~\ref{fig:ex171_traj}.

For $f_{dim} = (0,1)$, as the sail gets bigger, it performs better with a lower $\Delta t(N)$. This can be explained by a combination of two factors:
\begin{enumerate}
\item Higher $|\mathbf{a}_{SRP}|$ of larger sails:\\
If $|\mathbf{a}_{SRP}|$ is too big compared to the gravitational acceleration of the central body, if a solar sail maintains the same attitude for a longer time (larger $\Delta t(N)$), the sail behaves as a perturbation source by deviating the trajectory too much from the goal. In this case, lower $\Delta t(N)$ have a better chance of achieving the desired trajectory.
\item Search strategy (Beam Search) employed by the agent:\\
a beam search is a form of greedy search. This means that its choices do not rely on past decisions. However, nodes with a larger $\Delta t(N)$ store more information from the past, which provides the agent with better information for its choices.
\end{enumerate}
As a consequence, a balance is achieved: solar sails should perform better for lower $\Delta t(N)$, but the agent is capable of finding better solutions for higher $\Delta t(N)$.

The larger asteroids considered, with $f_{dim} = (2,3)$, present a critical scenario to the employment of solar sails. Smaller $l_{boom}$ are no longer viable in some cases and, as better discussed in Section~\ref{sec:res/lan}, the agent has greater difficulty in maintaining a dusk-dawn configuration. 

\begin{figure}[!ht]
    \centering
    \includegraphics[width=0.6\columnwidth]{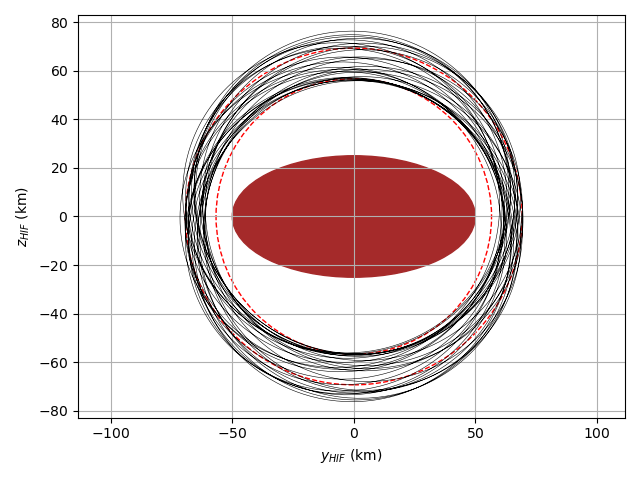}
    \caption{Trajectory for $f_{dim} = 2$, $l_{boom} = 3.0\,\mathrm{m}$ and $\Delta t(N) = 4.0 \times T_0$.}
    \label{fig:ex234_traj}
\end{figure}

For $f_{dim} = 2$, a difference in $\Delta t(N)$ makes little impact on results. The values of a $l_{boom}$, presented in Table~\ref{tab:results/cond1}, beat by a small margin results from the same $l_{boom}$. The trajectory of one of these cases is shown in Figure~\ref{fig:ex234_traj}. It is visually clear that the trajectory escapes the tolerance region considerably more than the trajectory from Figure~\ref{fig:ex171_traj}.

\begin{figure}[!ht]
    \centering
    \includegraphics[width=0.6\columnwidth]{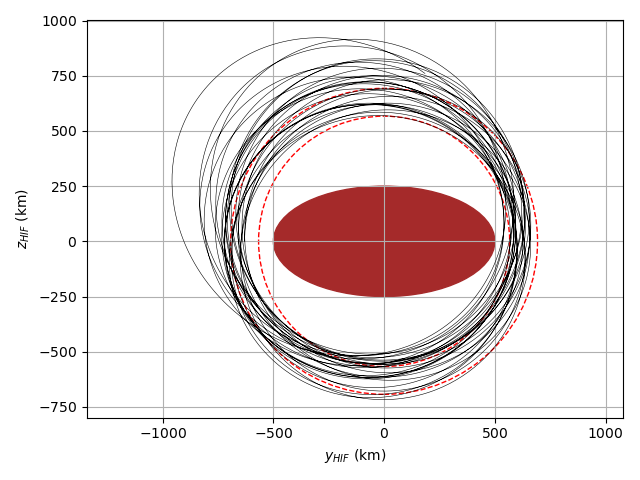}
    \caption{Trajectory for $f_{dim} = 3$, $l_{boom} = 5.0\,\mathrm{m}$ and $\Delta t(N) = 8.0 \times T_0$.}
    \label{fig:ex358_traj}
\end{figure}

At last, for $f_{dim} = 3$, only a few successful results were obtained by the agent, for larger $l_{boom}$ and $\Delta t(N)$. Nevertheless, these results presented lower $\eta$. The trajectory from $f_{dim} = 3$ and $l_{boom} = 5.0\,\mathrm{m}$ is shown in Figure~\ref{fig:ex358_traj} and should be considered as an outlier. Despite not resulting in a collision, the sail only acts as a means to delay it. The trajectory is similar to the one seen in Figure~\ref{fig:exOff_traj}.

\begin{figure}[!ht]
    \centering
    \includegraphics[width=0.6\columnwidth]{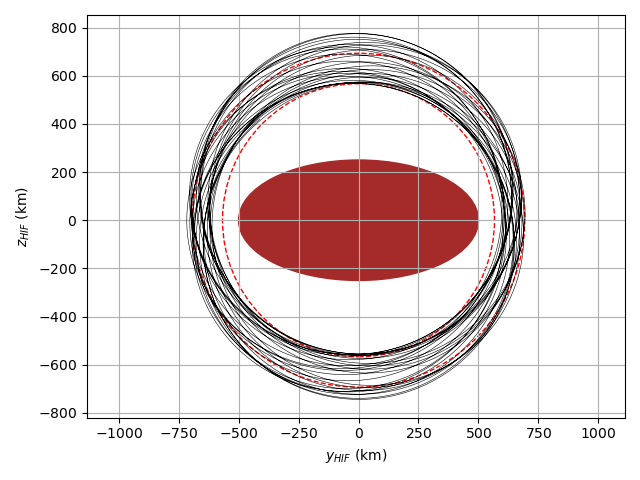}
    \caption{Trajectory for $f_{dim} = 3$, $l_{boom} = 7.0\,\mathrm{m}$ and $\Delta t(N) = 8.0 \times T_0$.}
    \label{fig:ex378_traj}
\end{figure}

The agent was able to sustain the trajectory for \added{$30 \times T_0$} with the largest sail, combined with the longest $\Delta t(N)$. Once again, this can be explained by the search strategy employed. The trajectory is illustrated in Figure~\ref{fig:ex378_traj}.

\subsection{Inclination maintenance}
\label{sec:res/inc}

An additional analysis can be made on the effectiveness of the agent in maintaining a polar orbit $(KE_i = 90^\circ \pm 1^\circ)$. Table~\ref{tab:results/cond2} displays results analogous to Table~\ref{tab:results/cond1}, but focused only in evaluating when $goal_i$ is respected.

\begin{table*}
\centering
\caption{Decision time $\Delta t(N)$, measured in $T_0$, with the largest $\eta$, in parenthesis, for $goal_i \rightarrow$ True}
\begin{tabular}{|c|*{4}{c|}}
\hline
\diagbox[]{$f_{dim}$}{$l_{boom}\,\mathrm(m)$} & $1.0$ & $3.0$ & $5.0$ & $7.0$ \\ 
\hline $0$
& $0.5\ (98.89\%)$ & $0.5\ (100.00\%)$ & $0.5\ (99.51\%)$ & $0.5\ (98.16\%)$ \\
\hline $1$
& $2.0\ (100.00\%)$ & $1.0\ (90.77\%)$ & $0.5\ (93.69\%)$ & $0.5\ (97.70\%)$ \\
\hline $2$
& N/A & $0.5\ (100.00\%)$ & $0.5\ (100.00\%)$ & $2.0\ (97.25\%)$  \\
\hline $3$
& N/A & N/A & $8.0\ (100.00\%)$ & $4.0\ (100.00\%)$  \\
\hline
\end{tabular}
\label{tab:results/cond2}
\end{table*}

By this analysis, it is clear that the smallest $\Delta t(N) = 0.5 \times T_0$ result in better inclination maintenance. Scenarios where this is not the case, either have a small difference of $\eta$, with $\Delta t(N) = 0.5 \times T_0$ losing by a small margin (e.g.: $f_{dim} = 1$, $l_{boom} = 3.0\,\mathrm{m}$ ; $f_{dim} = 2$, $l_{boom} = 7.0\,\mathrm{m}$), or smaller $\Delta t(N)$ resulted in collisions (e.g.: $f_{dim} = 1$, $l_{boom} = 1.0\,\mathrm{m}$ ; $f_{dim} = 3$, $l_{boom} = 5.0\,\mathrm{m}$ ; $f_{dim} = 3$, $l_{boom} = 7.0\,\mathrm{m}$).

\subsection{Longitude of ascending node update}
\label{sec:res/lan}

Another possible analysis is the agent capacity of maintaining a dusk-dawn configuration. It should update the longitude of ascending node according to the movement of the asteroid in its own heliocentric orbit. Once again, the quality of maintenance is presented in Table~\ref{tab:results/cond3} but, this time, focused only on how long $goal_\Omega$ was respected.

\begin{table*}
\centering
\caption{Decision time $\Delta t(N)$, measured in $T_0$, with the largest $\eta$, in parenthesis, for $goal_\Omega \rightarrow$ True}
\begin{tabular}{|c|*{4}{c|}}
\hline
\diagbox[]{$f_{dim}$}{$l_{boom}\,\mathrm(m)$} & $1.0$ & $3.0$ & $5.0$ & $7.0$ \\ 
\hline $0$
& $0.5\ (85.43\%)$ & $2.0\ (93.03\%)$ & $0.5\ (96.64\%)$ & $0.5\ (97.16\%)$ \\
\hline $1$
& $2.0\ (9.78\%)$ & $1.0\ (90.13\%)$ & $2.0\ (87.35\%)$ & $0.5\ (90.18\%)$ \\
\hline $2$
& N/A & $0.5\ (11.95\%)$ & $4.0\ (25.24\%)$ & $4.0\ (80.67\%)$ \\
\hline $3$
& N/A & N/A & $8.0\ (9.33\%)$ & $4.0\ (9.59\%)$ \\
\hline
\end{tabular}
\label{tab:results/cond3}
\end{table*}

For $f_{dim} = (0,1)$, similar conclusions to Section~\ref{sec:res/inc} can be taken: lower $\Delta t(N)$ are better at updating $KE_\Omega$, all beating each other by small margins. Therefore, not always the lowest $\Delta t(N)$ corresponds to the best $\eta$, but they all presents similar results.

\begin{figure}[!ht]
    \centering
    \includegraphics[width=0.6\columnwidth]{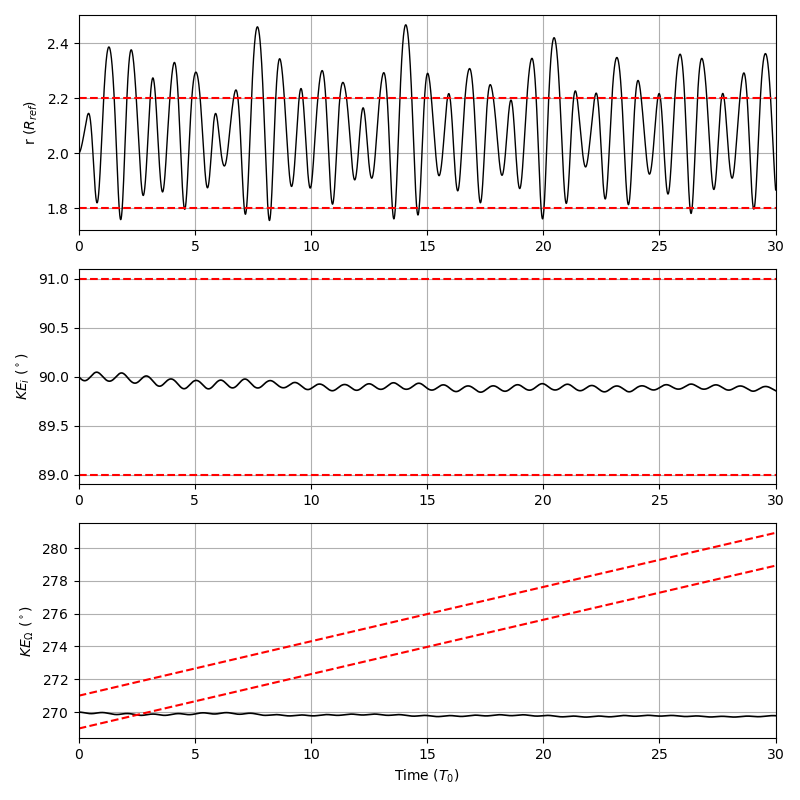}
    \caption{Heuristic components for $f_{dim} = 3$, $l_{boom} = 7.0\,\mathrm{m}$ and $\Delta t(N) = 8.0 \times T_0$.}
    \label{fig:ex378_keps}
\end{figure}

However, $f_{dim} = (2,3)$ present a different scenario. Results with $\eta$ that are far too low (smaller than $10\%$) indicate that the sail is not capable of updating $KE_\Omega$. One of this cases is illustrated in Figure~\ref{fig:ex378_keps}, where it is clear $KE_\Omega$ remains constant and does not follow the movement of the asteroid in its heliocentric orbit. There is a tendency to, little by little, shift from a dusk-dawn configuration to a noon-midnight one. In this case, an orbital maneuver will be necessary to perform a correction of $KE_\Omega$ on a future instant of the mission. 

\subsection{Station keeping}
\label{sec:res/stationKeep}

Finally, Table~\ref{tab:results/cond0} presents an overall evaluation of how effective the agent is in respecting all the goals together. Therefore, it indicates for how long the agent was able to maintain a circular and polar orbit, with a dusk-dawn configuration, within a $10\%$ distance to the initial distance $r_0$.

\begin{table*}
\centering
\caption{Decision time $\Delta t(N)$, measured in $T_0$, with the largest $\eta$, in parenthesis, for all$(goals) \rightarrow$ True}
\begin{tabular}{|c|*{4}{c|}}
\hline
\diagbox[]{$f_{dim}$}{$l_{boom}\,\mathrm(m)$} & $1.0$ & $3.0$ & $5.0$ & $7.0$ \\ 
\hline $0$
& $2.0\ (48.10\%)$ & $2.0\ (59.87\%)$ & $1.0\ (61.87\%)$ & $0.5\ (58.40\%)$ \\
\hline $1$
& $2.0\ (6.41\%)$ & $2.0\ (50.63\%)$ & $2.0\ (55.44\%)$ & $1.0\ (62.78\%)$ \\
\hline $2$
& N/A & $0.5\ (7.68\%)$ & $4.0\ (17.46\%)$ & $2.0\ (45.40\%)$ \\
\hline $3$
& N/A & N/A & $8.0\ (6.00\%)$ & $4.0\ (6.24\%)$ \\
\hline
\end{tabular}
\label{tab:results/cond0}
\end{table*}

\begin{figure}[!ht]
    \centering
    \includegraphics[width=0.6\columnwidth]{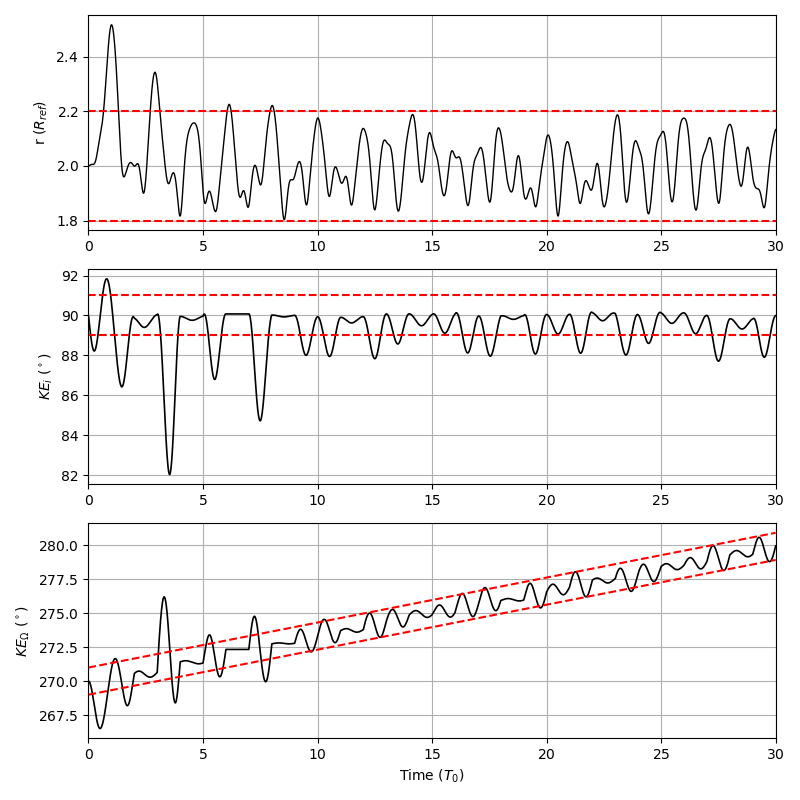}
    \caption{Heuristic components for $f_{dim} = 1$, $l_{boom} = 7.0\,\mathrm{m}$ and $\Delta t(N) = 1.0 \times T_0$.}
    \label{fig:ex171_keps}
\end{figure}

Additionally, the values of $r$, $KE_i$ and $KE_\Omega$ for the best solution (largest $\eta$), are presented in Figure~\ref{fig:ex171_keps}.

\begin{figure}[!ht]
    \centering
    \includegraphics[width=0.6\columnwidth]{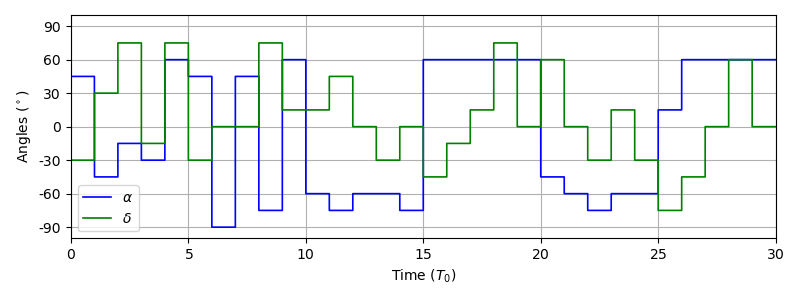}
    \caption{Time history of the attitude angles for $f_{dim} = 1$, $l_{boom} = 7.0\,\mathrm{m}$ and $\Delta t(N) = 1.0 \times T_0$.}
    \label{fig:ex171_angs}
\end{figure}

As an example of the results obtained by the agent, the time history of the attitude angles $(\alpha,\delta)$ are presented in Figure~\ref{fig:ex171_angs}.

Even though the focus of this study is not solely the time history of $(\alpha,\delta)$, it is possible to present a brief discussion of these results. From Figure~\ref{fig:ex171_angs}, it is readily seen that $\Delta t(N) = 1.0 \times T_0$, given the changes of $\alpha$ and $\delta$ once every $1.0 \times T_0$ time interval. Until $t = 10.0 \times T_0$, there are more abrupt attitude changes, specially considering the variations of $\alpha$. After $t = 10.0 \times T_0$, the attitude changes are smoother and $\alpha$ has major changes practically once every $5.0 \times T_0$. From this analysis and results from Figure~\ref{fig:ex171_keps}, is possible to affirm that, in this case, the agent was capable of finding a steady long-term operational regime for its station-keeping mission.

\added{The agent assumed instantaneous changes of the spacecraft attitude while calculating the solutions. However, real-life scenarios require a finite amount of time for attitude changes. By limiting the number of attitude changes to a minimum of $\Delta t(N) = 0.5 \times T_0 \approx 3h\,18min$, this concern is partly addressed. This would allow the use of lower torque attitude control systems, such as Reflectivity Control Devices (RCDs)~\citep{funase:2010}. However, these results are sensitive to non-instantaneous attitude changes and this is a limitation of the trajectories calculated in this study.}

Overall results for $f_{dim} = (2,3)$ are compromised by the sail's inability to update $KE_\Omega$. For this conditions, there also are cases (lower $l_{boom}$) where the spacecraft collides with the asteroid, in spite of the solar sail.

It is possible to suggest suitable sail sizes for the different asteroid sizes considered, as presented in Table~\ref{tab:results/sizes} with a check mark $(\checkmark)$.

\begin{table*}
\centering
\caption{Suitable sail sizes for the different asteroid dimensions}
\begin{tabular}{|c|*{4}{c|}}
\hline
\diagbox[]{$f_{dim}$}{$l_{boom}\,\mathrm(m)$} & $1.0$ & $3.0$ & $5.0$ & $7.0$ \\ 
\hline $0$
& $\checkmark$ & $\checkmark$ & $\checkmark$ & $\checkmark$ \\
\hline $1$
& $-$ & $\checkmark$ & $\checkmark$ & $\checkmark$ \\
\hline $2$
& $-$ & $-$ & $-$ & $\checkmark$ \\
\hline $3$
& $-$ & $-$ & $-$ & $-$ \\
\hline
\end{tabular}
\label{tab:results/sizes}
\end{table*}

Within the suggested sizes, the agent is capable of respecting all of the imposed conditions in between $45\%$ to $63\%$ of the time. It is important to note that, for example, if the tolerances were relaxed to $\mathbf{tol} = [15\% \cdot r_0 , 2^\circ , 2^\circ]$, than all the $\eta$ values of these cases would range from $76\%$ to $94\%$.

For $(f_{dim} = 0\,; l_{boom} = 7.0\,\mathrm{m})$, the largest $\eta$ occurred for $\Delta t(N) = 0.5 \times T_0$. For $(f_{dim} = 0\,; l_{boom} = 5.0\,\mathrm{m})$ and $(f_{dim} = 1\,; l_{boom} = 7.0\,\mathrm{m})$, the best alternative is $\Delta t(N) = 1.0 \times T_0$. The remainder of the listed cases have better results for $\Delta t(N) = 2.0 \times T_0$.

This means that, on some of these scenarios, it is viable to station-keep a spacecraft over an irregularly shaped asteroid with one attitude change per, slightly over, $13$~hours. Additionally, if the asteroid is small enough, this can be achieved with sails smaller than the ones built in past missions. This is an advantage both from a structural point of view - an easier to build sail - and an attitude control system implementation - a smaller sail, with fewer attitude maneuvers, requires less effort from the attitude actuators.

Finally, the resulting rules of thumb can be determined:
\begin{enumerate}
    \item Larger sails require smaller $\Delta t(N)$;
    \item Bigger asteroids allow for larger $\Delta t(N)$, for the same $l_{boom}$;
    \item Larger asteroid require larger sails.
\end{enumerate}

The initial orbital conditions analyzed in this study resulted in a trajectory where the spacecraft collided with the asteroid without the use of a solar sail. But another concern would be an early escape from the asteroid gravitational field of the asteroid. This would be a growing concern for initial orbits which are farther away from the asteroid, such as $KE_{a}(t_0) = 3.0 \times R_{ref}$. In this case, larger sails would have inferior performance, specially for smaller asteroids. This would be an interesting topic to explore in future studies.

\subsection{Acceleration quotient}
\label{sec:res/acc}

It is interesting to make a brief analysis of an acceleration quotient $\left( Q = \frac{|\mathbf{a}_{SRP}|}{|\mathbf{a}_{g,ast\rightarrow sc}|} \right)$ between all the different scenarios. Bigger asteroids produce larger $|\mathbf{a}_{g,ast\rightarrow sc}|$, and larger sails provide bigger $|\mathbf{a}_{SRP}|$. The initial orbital conditions are used to calculate the magnitude of these accelerations, with an additional consideration that the sail is completely exposed to the sunlight $(\,\alpha = 0^\circ\,,\,\delta = 0^\circ\,)$.

\added{From Equations~\ref{eq:accSRP} and~\ref{eq:light}, $|\mathbf{a}_{SRP}| \propto S_{sail}$ when all the other conditions are maintained. On the other hand, $S_{sail} \propto l_{boom}^2$ and $|\mathbf{a}_{SRP}| \propto l_{boom}^2$. Alternatively, from Equation~\ref{eq:accMB}, $|\mathbf{a}_{g,ast\rightarrow sc}| \propto \frac{\mu_{ast}}{r^2_{ast \rightarrow sc}}$. In addition, from the definition of the volume of an ellipsoid and an asteroid with constant density $\mu_{ast} \propto 10^{3f_{dim}}$ and from an initial distance proportional to the reference radius of the asteroid $r_{ast \rightarrow sc} \propto 10^{f_{dim}}$. Therefore, $|\mathbf{a}_{g,ast\rightarrow sc}| \propto 10^{f_{dim}}$ and, with the same sail size, the acceleration quotient is multiplied by $10^{-f_{dim}}$ when considering different asteroids.}

\added{Finally, by varying the sail sizes with $l_{boom}$ and the asteroid sizes and initial orbital conditions with $f_{dim}$ and maintaining all other parameters constant, it is possible to deduce that $Q \propto l_{boom}^2 \times 10^{-f_{dim}}$. This relation is confirmed by calculating $Q$ for the $16$~different combinations of $(f_{dim},l_{boom})$ and initial conditions considered. The results can be presented in the form of the empirically obtained equation:}

\added{
\begin{equation}
Q \approx (2.35)\,l_{boom}^2 \times 10^{(-2)-f_{dim}}
\label{eq:accQuot}
\end{equation}
}

From \added{Equation~\ref{eq:accQuot}} it is possible to verify that all the collision scenarios, alongside the outlier $(f_{dim} = 3\,;l_{boom} = 5.0\,\mathrm{m})$, $Q < 1.0 \times 10^{-3}$. This means that $|\mathbf{a}_{SRP}|$ is too small compared to $|\mathbf{a}_{g,ast\rightarrow sc}|$. For scenarios with low $\eta$ (From Table~\ref{tab:results/cond0}), where the agent was unable to update $KE_\Omega$, $1.0 \times 10^{-3} < Q < 1.0 \times 10^{-2}$. This could be interpreted as a boundary region where the sail operates near its environment limitations. Finally, the successful cases present $Q > 1.0 \times 10^{-2}$. These are the regions where the sail is large enough for station keeping.

\section{Conclusions}
\label{sec:conc}

This study \added{compared the use of different sized solar sails for station keeping a spacecraft over asteroids with different sizes. A} rational agent \added{was used} as an attitude guidance system for \added{the} solar sails \added{and the asteroids were modeled with irregular shapes.} The agent had the goal of maintaining the spacecraft in a polar orbit with a dusk-dawn configuration, within $10\%$ of the initial distance. An initial low altitude scenario was considered, where the spacecraft rapidly collided with the asteroid if no orbital maneuvers were performed. In addition, different attitude change rates were considered with different sizes of sails and asteroids \added{in an effort to find solutions with fewer amount of attitude changes throughout the mission.} The \added{use of the rational agent} allows a fast formulation of the problem which, in turn, enables the analysis of multiple scenarios. \added{The obtained results showed that this new approach served as a useful and valid alternative to studies of this nature.}

Overall, fewer attitude changes throughout a mission and smaller sails translate to a smaller effort from the attitude control system. For the same sail size, a larger asteroid allowed for a mission with fewer attitude changes. However, there is a threshold where the sail is too small to produce enough acceleration to maintain the orbit. In this case, larger sails should be used. However, larger sails require more attitude changes. An acceleration quotient \added{$(Q)$} was proposed to rapidly define if the sail is large enough to station-keep at the specified asteroid. If the solar radiation pressure acceleration from the sail is larger than $1\%$ of the gravitational acceleration from the asteroid, than it is possible to station-keep at the explored conditions. 

\added{The identified $Q$ ranges for feasible station keeping are taken from a discrete set of simulations and were empirically established. Additionally, the results are mainly influenced by the search technique employed by the agent. If the agent search parameters are well adjusted by a careful set of test simulations, they present a minor role in the search process.} Further studies can investigate if this relation is sustained for asteroids with different shapes and different orbital conditions.

\section*{Acknowledgments}
The authors wish to express their appreciation for the support provided by the Coordination for the Improvement of Higher Education Personnel (CAPES – code 001), the National Council for Scientific and Technological Development (CNPq), grants \#175425/2023-0, \#316991/2023-6 and \#309089/2021-2 and the São Paulo Research Foundation (FAPESP), grant \#2022/11783-5. This research was supported by resources supplied by the Center for Scientific Computing (NCC/GridUNESP) of the São Paulo State University (UNESP).

\bibliographystyle{plainnat}
\bibliography{./bib/ref_SolarSails,./bib/ref_SailMissions,./bib/ref_OrbMec,./bib/ref_Optimization,./bib/refs}

\end{document}